\documentclass[%
 amsmath,amssymb,
preprint,%
]{revtex4-1}
\usepackage{graphicx}
\usepackage{dcolumn}
\usepackage{bm}
\usepackage{color}
\usepackage{epstopdf}

\begin{document}
\title{Magnetic-field-induced crossover from the inverse Faraday
effect \\ to the optical orientation in EuTe}
\author{V.~V.~Pavlov}
\affiliation{Ioffe Institute, the Russian Academy of Sciences, 194021 St. Petersburg, Russia}
\author{R.~V.~Pisarev}
\affiliation{Ioffe Institute, the Russian Academy of Sciences, 194021 St. Petersburg, Russia}
\author{S.~G.~Nefedov}
\affiliation{Ioffe Institute, the Russian Academy of Sciences, 194021 St. Petersburg, Russia}
\author{I.~A.~Akimov}
\affiliation{Ioffe Institute, the Russian Academy of Sciences, 194021 St. Petersburg, Russia}
\affiliation{Experimentelle Physik 2, Technische Universit\"at Dortmund, 44221 Dortmund, Germany}
\author{D.~R.~Yakovlev}
\affiliation{Ioffe Institute, the Russian Academy of Sciences, 194021 St. Petersburg, Russia}
\affiliation{Experimentelle Physik 2, Technische Universit\"at Dortmund, 44221 Dortmund, Germany}
\author{M.~Bayer}
\affiliation{Ioffe Institute, the Russian Academy of Sciences, 194021 St. Petersburg, Russia}
\affiliation{Experimentelle Physik 2, Technische Universit\"at Dortmund, 44221 Dortmund, Germany}
\author{A.~B.~Henriques}
\affiliation{Instituto de F\'{\i}sica, Universidade de S\~{a}o Paulo, 05315-970 S\~{a}o Paulo, Brazil}
\author{P. H. O. Rappl}
\affiliation{LAS-INPE, 12227-010 Sao Jose dos Campos, Brazil}
\author{E. Abramof}
\affiliation{LAS-INPE, 12227-010 Sao Jose dos Campos, Brazil}
\date{\today }

\begin{abstract}
A time-resolved optical pump-probe technique has been applied for studying the
ultrafast dynamics in the magnetic semiconductor EuTe near the absorption
band gap. We show that application
of external magnetic field up to 6~T results in crossover
from the inverse Faraday effect taking place on the femtosecond time scale to the optical
orientation phenomenon with an evolution in the picosecond time domain. We propose a model which includes both these processes possessing different
spectral and temporal properties. The circularly polarized optical pumping induces the optical
electronic transition $4f^75d^0 \rightarrow 4f^65d^1$ forming the absorption band gap in EuTe.
The observed crossover is related to a strong
magnetic-field shift of the band gap in EuTe at low temperatures.
It was found that manipulation of spin states on intrinsic
defect levels takes place on a time scale of 19~ps in the applied magnetic field of 6~T.
\end{abstract}
\pacs{78.47.D-, 75.50.Pp, 75.50.Ee}
\maketitle

\section{INTRODUCTION}
Ultrafast phenomena related to the control and manipulation of electronic and spin states are in the focus of current research in
physics of magnetism, magnonics and spintronics \cite{Magnetism, Magnonics, Spintronics}. This activity is motivated, first of all, by novel experiments on ultrafast magnetic phenomena revealing new fundamentally important mechanisms of electronic-spin dynamics taking place on femto- and picosecond time scales \cite{Kalashnikova,Kimel, Pavlov1, Kirilyuk1, Bigot, Chen}. On the other hand, new results open potential possibilities for constructing high-speed magneto-electronic and magneto-optical devices.

The conservation of angular momentum is one of the most
fundamental law of physics which plays important role in various phenomena. For example, the Einstein-de Haas effect is a consequence of this conservation manifesting in the mechanical rotation of a free body, when its magnetic moment is changed \cite{Einstein}. Photon is a particle with an intrinsic angular momentum of one unit of $\hbar$ in the quantum-mechanical description, therefore the circularly polarized light carries a spin angular momentum. In Ref.~\cite{Allen} it is shown that the torque exerted by circularly polarized light can be transferred to a small electric dipole. An intense circularly polarized light may create a magnetization $\mathbf{M}$ in a medium during the photon-electron interaction due the inverse Faraday effect:
\begin{equation}
\mathbf{M}(0)=-i\mathbf{\chi^{(2)}}(0;\omega,-\omega) \left[\mathbf{E}(\omega) \times \mathbf{E}^{\ast }(\omega)\right],
\label{Eq1}
\end{equation}
where $\mathbf{\chi^{(2)}}$ is the second order nonlinear optical susceptibility describing the two-photon mixing process allowed in any media~\cite{Shen}, $\mathbf{E}$ is an oscillating electric field at angular frequency $\omega$. This phenomenon was theoretically predicted and discussed in Refs.~\cite{Pitaevskii,Pershan0,Pershan}, and experimentally observed in Ref.~\cite{Ziel}. Nowadays, the inverse Faraday effect is widely used for experimental and theoretical studies of ultrafast phenomena in magnetic systems~\cite{Kirilyuk1, Berritta, Berritta1, Qaiumzadeh, Freimuth}. An inverse transverse magneto-optical Kerr effect related to Eq.~(\ref{Eq1}) was predicted in Ref.~\cite{Belotelov}.

Another physical phenomenon based on the angular momentum transfer from circularly polarized light to a medium is the optical orientation.
This phenomenon reflects the exchange of angular momentum between the circularly polarized light and atomic or solid state systems. The principles of optical orientation were established by A. Kastler for paramagnetic atoms \cite{Kastler}, then were successfully applied for molecules \cite{Auzinsh} and semiconductors \cite{Orient}. For example, due to the angular momentum conservation a circularly polarised photon creates a spin-oriented $s$-electron in GaAs with a rather long life time at room \cite{Kimel1} and low temperatures \cite{Belykh}. The phenomenon of optical orientation is a linear optical process taking place during the interaction of circularly polarized light with an absorbing medium.

Here we report on the magnetic-field control of interplay between the inverse Faraday effect and optical orientation close to the absorption band gap in the magnetic semiconductor EuTe. We show that mechanisms
of optical orientation and spin-relaxation in these materials are different from those in model band semiconductors $A^{III}B^{V}$ and $A^{II}B^{VI}$
due to a specific electronic structure of EuTe.

\section{EXPERIMENTAL DETAILS AND RESULTS}
Magnetic semiconductors Eu\textit{X}(\emph{X}= O, S, Se, Te) represent a group of materials possessing unique
electronic, magnetic, optical, and magneto-optical properties~\cite{Wachter,Gunther,Nagaev} which are determined by strongly localized 4\emph{f}$^7$
electrons of the Eu$^{2+}$ ions with spin $S=7/2$ and orbital moment $L=0$. EuTe is antiferromagnetic with a N\'{e}el
temperature $T_N = 9.6$~K. The magnetic moments of the two sublattices $\mathbf{m}_1$ and $\mathbf{m}_2$ ($|m_1|=|m_2|$) are ordered antiferromagnetically in adjacent (111)-planes. In external magnetic field the EuTe can be ferromagnetically saturated аbove a critical field $B_c = 7.2 T$ .
Most of the early research in 1960s-1970s were performed on bulk single crystals and
noncrystalline thin films of Eu\textit{X}. However, during last decade high-quality epitaxial thin films of Eu\textit{X} were successfully grown on
Si and GaN semiconductor substrates opening new opportunities for applications \cite{Lettieri,Schmehl,Averyanov}. Eu\textit{X} compounds reveal a new type of
nonlinear magneto-optical effects \cite{Kaminski,Lafrentz}, ultrafast spin dynamics \cite{APL2010, PRL2012, SciRep2014, NatureComm2015}, photo-induced spin polarons with a giant magnetic moment \cite{Henriques1,Henriques2}.
\begin{figure}[]
\includegraphics[width=0.45\textwidth]{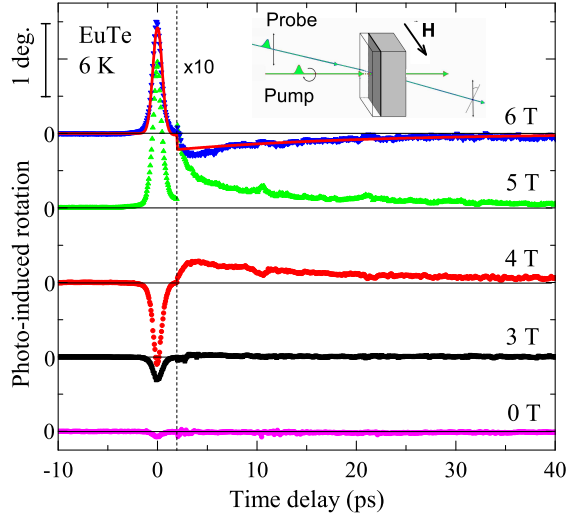}
\caption{(color online). Temporal behavior of the photo-induced
optical rotation in EuTe for different magnetic fields at the photon energy of 2.19 eV.
For clarity the traces are shifted vertically. A solid curve for magnetic field of 6~T shows the best fit on the basis of Eq.~(\ref{photo}) for
the following parameters $A=1.69(1)^\circ$, $B=-0.024(3)^\circ$, $\sigma=0.67(1)$~ps and $\tau_s=19(3)$~ps. There is a scale change of ordinate axis by enhance factor of 10 at the time delay of 2ps (noted as x10).
Inset shows schematically the pump-probe experimental geometry.} \label{fig:figure1}
\end{figure}

We present results on optical pump-probe studies of epitaxial films of the
magnetic semiconductor EuTe. This material exhibits a very strong redshift of the fundamental absorption
edge by 130~meV for magnetic fields between 0 and 8~T \cite{Heiss1}. Optical effects for photon energies close to the EuTe absorption band gap could be governed by applying external magnetic field. Using a pump-probe technique, we performed experiments on the magnetic-field control of helicity-dependent photo-induced phenomena in EuTe.
Pump-probe experiments were done in transmission geometry using an optical parametric oscillator pumped by a Ti:Sapphire laser with 1~ps pulses at 80~MHz repetition rate. We used a degenerate optical scheme for pump and probe beams having photon energy of 2.19~eV. This energy is slightly below the band gap value of 2.4~eV in EuTe at zero magnetic field. EuTe films were grown by the molecular-beam epitaxy on (111)-oriented BaF$_2$ substrates \cite{Henriques3,Heiss}. The 1 $\mu$m thick layers were capped with a 40-nm-thick BaF$_2$ protective
layer and the high sample quality was confirmed by x-ray
analysis.

Figure 1 shows the probe light polarisation rotation induced by the pump beam (photo-induced
rotation) in EuTe as a function of the pump-probe time delay for different magnetic fields.
Magnetic fields up to 6~T were applied in the Voigt geometry $\mathbf{k}\parallel (111) \perp \mathbf{H}$, see Inset in Fig.~1.
Temporal behavior of the photo-induced
rotation is
characterized by a narrow Gaussian-shape peak around the zero time delay for magnetic fields of 0-3~T. For magnetic fields above 3~T a broad tail with a characteristic relaxation time of several picoseconds begins to appear.
\begin{figure}[]
\includegraphics[width=0.40\textwidth]{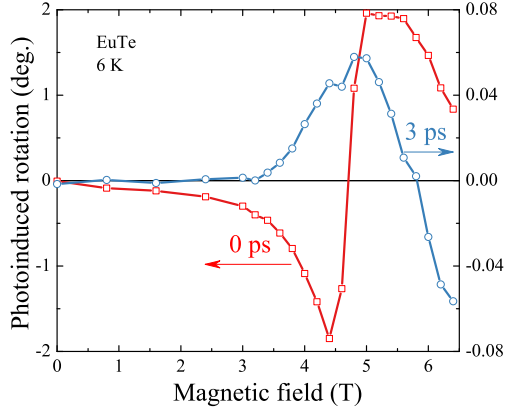}
\caption{(color online). Magnetic field dependence of the
photo-induced rotation in EuTe for two time delays at the photon energy of 2.19 eV.}
\label{fig:figure2}
\end{figure}

Figure 2 shows the photo-induced optical rotation as a function of magnetic field for two time delay of 0 and 3~ps. These two dependencies display appreciably distinct behavior. The magnetic field dependence for 0~ps time delay has nonzero values for magnetic fields $0<H<4.7$~T; it has a minimum at $H\simeq4.4$~T, then it reverses sign for magnetic field of 4.7~T and has a broad maximum for $H\simeq5.3$~T.  Surprisingly, the magnetic field dependence for 3~ps time delay has zero values for magnetic fields $0<H<3.2$~T. It has a maximum at $\simeq4.9$~T and reverses sign for $H\simeq5.8$~T. Quantitatively, the sign reversal of the photo-induced optical rotation for 0~ps is related to the strong redshift of the absorption band gap in EuTe in external magnetic field~\cite{Heiss1} when the fundamental absorption edge crosses the probe photon energy of 2.19~eV. Observed different behavior for time delays of 0 and 3~ps can be qualitatively understood taking into account an influence of excited electronic states on the absorption band gap in EuTe.

\begin{figure}[]
\includegraphics[width=0.4\textwidth]{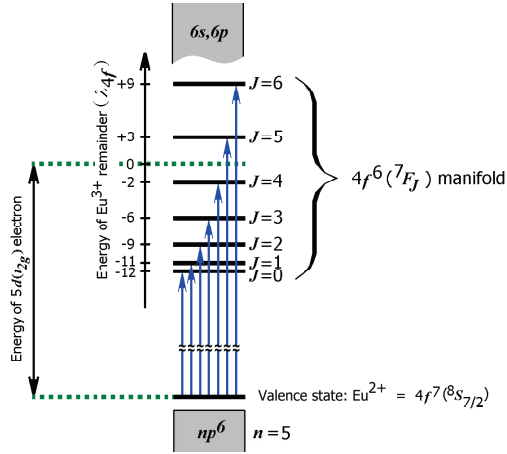}
\caption{(color online). Electronic energy diagram for
unpertubed $4f^7$ and excited $4f^6$ states in EuTe.} \label{fig:figure2}
\end{figure}
\section{DISCUSSION}
 For explaining optical pump-probe experiments let as consider the electronic energy diagram of the magnetic semiconductor EuTe in which Eu$^{2+}$ ions in the ground state $4f^7 5d^0$ play the decisive role. The electric dipole transition selection rules imply that the $4f \rightarrow 6s$ transitions are forbidden and  one has to take into account the $4f^7 5d^0 \rightarrow 4f^6 5d^1$ electric dipole transition forming the absorption band gap in EuTe \cite{Gunther1}.
Figure~3 shows the electronic energy diagram for unpertubed $4f^7$ and excited $4f^6$ states. The absorption edge of EuTe corresponds to the onset of the $4f^7 (^8S_{7/2})\rightarrow 4f^6 (^7F_J) 5d(t_{2g})$ transition, where the final state is combined of one $5d$-electron and six $4f$-electrons. The $^7F_J$-multiplet ($J = 0, 1 ...6$) of the six $4f$-electrons with a total spin-orbital splitting is about 0.6~eV. In magnetic fields of ~0.5~T the semiconductor EuTe has two antiferromagnetically ordered sublattices with spins oriented perpendicularly to the magnetic field $\textbf{H}$. Due to spin conservation in the electric-dipole $4f^7 5d^0\rightarrow 4f^6 5d^1$ absorption process for the circularly polarized photon, immediately after excitation, the electron spin is oriented along the spin of a magnetic sublattice. Moreover, following the Franck-Condon principle, the electronic transition takes place at fixed spatial and spin coordinates of the lattice. Schematically this process, which we call as Stage I, is shown in Fig. 4. The width of a single sublevel of the $^7F_J$-multiplet is about 0.1~eV which corresponds to the electron lifetime of about 18~fs. After this time interval, some electrons are trapped by a long-living intrinsic
defect states $4f^6 X^1$ (Stage II) which are responsible for the luminescence process \cite{Heiss1} and magnetic polaron states \cite{Henriques1,Henriques2}. The life time of electrons on these states is longer than 1~ns depending on an applied magnetic field. During this time interval the spin of electron at the states $4f^6 X^1$ starts to precess (Stage III) due to the presence of external magnetic field $\mathbf{H}$. This precession can be analysed in terms of the Landau-Lifshitz equation with the Gilbert damping \cite{Landau,Gilbert,Mondal}:
\begin{equation}
\frac{\partial \mathbf{M}}{\partial t}=-\gamma\mathbf{M\times H}%
^{eff}+\alpha \mathbf{M\times }\frac{\partial \mathbf{M}}{\partial t},
\label{LLG}
\end{equation}
where $\mathbf{M}$ is the local magnetization, $\gamma$ is the gyromagnetic ratio, $\mathbf{H}^{eff}$ is the effective
magnetic field, which accounts the external field $\mathbf{H}$ and exchange field in EuTe, and $\alpha$ is the Gilbert damping constant.
In Ref.~\cite{Mondal} it was shown that the relativistic extrinsic spin-orbit coupling give rise
to a dominant local spin relaxation mechanism in magnetic solids. In the Voigt geometry $\mathbf{k}\parallel (111) \perp \mathbf{H}$, this spin-relaxation mechanism  corresponds to the transverse spin relaxation (see Fig.~4). Finally, trapped electrons at the $4f^6 X^1$ states relax to the $4f^7 (^8S_{7/2})$ states (Stage IV).

In order to estimate the characteristic spin relaxation time of electrons on the $4f^6 X^1$ states we propose the following scheme. Assuming that the pump and probe pulses have the Gaussian temporal behavior, the photo-induced rotation $\theta$ can be analyzed by a single-time
relaxation model \cite{Pavlov1}:
\begin{eqnarray}
\theta &=& \frac{A}{\sigma \sqrt{\pi}}\exp \left( -\frac{t^{2}}{\sigma^{2}}\right)
\nonumber \\
&+&\frac{B}{2}\exp \left( \frac{\sigma ^{2}}{4\tau_s ^{2}}-\frac{t}{\tau_s
}\right) \left[ 1-\texttt{erf} \left( \frac{\sigma }{2\tau_s
}-\frac{t}{\sigma }\right) \right], \label{photo}
\end{eqnarray}
where $t$ is the pump-probe time delay, $\sigma$ is the width of
the Gaussian pulse, $A$ and $B$ are coefficients related to the instantaneous Gaussian-type and non-instantaneous delayed terms, respectively. The first and second terms in Eq.~(\ref{photo})
describe instantaneous contribution due the inverse Faraday
effect and non-instantaneous spin-related
contribution, respectively. When the laser pulse duration is much shorter than thermal relaxation times, the instantaneous contribution can be analysed in terms of the third-order optical nonlinearity $\mathbf{\chi^{(3)}}$ characterising a four-wave mixing process which is relevant to the pump-probe experiment. Photo-induced polarization $\mathbf{P}(\omega)$ arising in this four-wave mixing process can be written as:
\begin{equation}
\mathbf{P}(\omega)\propto \mathbf{\chi^{(3)}}(\omega;\omega,-\omega,\omega)\colon \mathbf{E}^{pump}(\omega)\mathbf{E}^{pump}(\omega)^{*}\mathbf{E}^{probe}(\omega),
\label{Eq2}
\end{equation}
where  $\mathbf{E}^{pump}(\omega)$ and $\mathbf{E}^{probe}(\omega)$ are the optical
electric fields of pump and probe beams, respectively.
Eq.~(\ref{Eq2}) describes a nonlinear optical process enabling an angular momentum transfer from the incoming circularly-polarized pump beam to linearly-polarized probe beam. The photo-induced polarization $\mathbf{P}(\omega)$ is a source for the outgoing probe light. This process is related to the orbital movement of electrons, therefore it is called the instantaneous orbital contribution \cite{Pavlov1}. In the case of EuTe the third-order optical nonlinearity $\mathbf{\chi^{(3)}}$ can be rather strong for the $4f^75d^0 \rightarrow 4f^65d^1$ electric-dipole optical transition. This gives rise to the photo-induced rotation observed in our experiments which is due to the inverse Faraday effect.
\begin{figure}[]
\includegraphics[width=0.4\textwidth]{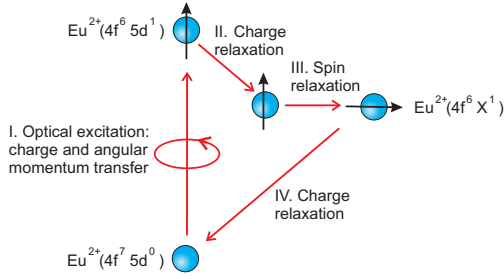}
\caption{(color online). Optical excitation (stage I) and relaxation processes (stages II, III, IV) at the dipole-allowed electronic transition
$4f^75d^0 \rightarrow 4f^65d^1$ in EuTe. } \label{fig:figure2}
\end{figure}

The non-instantaneous contribution appears for applied magnetic field values higher than 3~T. The absorption band gap of EuTe is about 2.4~eV at low temperatures, and this value is strongly influenced by an applied field \cite{Heiss1}. At magnetic field of 5~T the
band gap is about 2.2~eV. Thus, the non-instantaneous contribution is related to ultrafast optical orientation for the electric-dipole transition $4f^75d^0 \rightarrow 4f^65d^1$ with the subsequent charge and spin relaxation with a time constant $\tau_s$ (see Fig. 4). This relaxation corresponds for the Voigt geometry to the transverse component decay of the photo-induced magnetization vector $\mathbf{M}$ towards its equilibrium orientation parallel to the applied field $\mathbf{H}$. Applying fitting procedure on the basis of Eq.~(\ref{photo}) to the experimental data we found the transverse spin relaxation time $\tau_s=19$~ps at the $4f^6 X^1$ states. We note that this value is about two times shorter than precession period and about ten times shorter than decay time of precession oscillations at the $4f^7 d^0$ states in the applied magnetic field of 6~T~\cite{SciRep2014} .

\section{CONCLUSIONS}
In conclusion, we observed a strong optical response in the magnetic semiconductor EuTe for the circularly
polarized pump in transmission geometry applying magnetic field in
the Voigt geometry $\mathbf{k}\parallel (111) \perp \mathbf{H}$.
Observed signals can be attributed to the strong optical nonlinearity
of the third order accounting the inverse Faraday effect and the optical orientation phenomenon at electronic transition
from the localized $4f^7$ states of Eu$^{2+}$ ions on the top of the valence
band into $5d$ orbitals forming the conduction band.
Applying magnetic field in the range of 0-6~T one can control the interplay between the inverse Faraday effect and the optical orientation phenomenon in EuTe. We note that such crossover mechanism can be important for different classes of intrinsic and diluted magnetic semiconductors in which the band gap can be strongly influenced by the external magnetic field.

\section*{ACKNOWLEDGMENTS}
The authors are thankful to M.~M.~Glazov for useful discussions. This work was supported by the Russian Science Foundation (Grant No. 17-12-01314), the Dortmund group acknowledges support by the DFG (ICRC TRR 160), A.~B.~H. acknowledges support from the Brazilian agencies CNPq and FAPESP.


\begin{thebibliography}{[1]}
\bibitem{Magnetism} J. Walowski, M. M\"{u}nzenberg,  J. Appl. Phys. \textbf{120}, 140901 (2016).
\bibitem{Magnonics} V.~V.~Kruglyak, S.~O.~Demokritov, D.~Grundler, J. Phys. D: Appl. Phys. \textbf{43}, 260301 (2010).
\bibitem{Spintronics} \emph{Spintronics for Next Generation Innovative Devices}, eds. K. Sato, E. Saitoh, (Wiley,  Chichester, 2015).
\bibitem{Kalashnikova} A. M. Kalashnikova, A. V. Kimel, R. V. Pisarev, Physics Uspekhi \textbf{58}, 969 (2015).
\bibitem{Kimel} A. V. Kimel, A. Kirilyuk, P. A. Usachev, R. V. Pisarev, A. M. Balbashov, Th. Rasing, Nature \textbf{435}, 655 (2005).
\bibitem{Pavlov1} V. V. Pavlov, R. V. Pisarev, V. N. Gridnev, et. al., Phys. Rev. Lett. \textbf{98}, 047403
(2007).
\bibitem{Kirilyuk1} A. Kirilyuk, A. V. Kimel, T. Rasing, Rev. Mod. Phys. \textbf{82}, 2731
(2010).
\bibitem{Bigot} J.-Y. Bigot, M. Vomir, E. Beaurepaire, Nature Physics \textbf{5}, 515 (2009).
\bibitem{Chen} X.-J. Chen, Scient. Reports \textbf{7}, 41294 (2017).
\bibitem{Einstein} A. Einstein, W. J. de Haas, Koninklijke Akademie van Wetenschappen te Amsterdam, Proceedings \textbf{18}, 696 (1915).
\bibitem{Allen} P. J. Allen, Am. J. Phys. \textbf{34}, 1185 (1966).
\bibitem{Shen}Y.~R.~Shen, \emph{The Principles of Nonlinear Optics} (Wiley, New York, 1984).
\bibitem{Pitaevskii} L. P. Pitaevskii, Sov. Phys. JETP \textbf{12}, 1008 (1961).
\bibitem{Pershan0} P. S. Pershan, Phys. Rev. \textbf{130}, 919 (1963).
\bibitem{Pershan} P. S. Pershan, J. P. van der Ziel, L. D. Malmstrom, Phys. Rev. \textbf{143}, 574–583 (1966).
\bibitem{Ziel} J. P. van der Ziel, P. S. Pershan, L. D. Malmstrom, Phys. Rev. Lett. \textbf{15}, 190 (1965).
\bibitem{Berritta1} M. Battiato, G. Barbalinardo, P. M. Oppeneer, Phys. Rev. B \textbf{89} 014413 (2014).
\bibitem{Berritta} M. Berritta, R. Mondal, K. Carva, P. M. Oppeneer, Phys. Rev. Lett. \textbf{117}, 137203 (2016).
\bibitem{Qaiumzadeh} A. Qaiumzadeh, M. Titov, Phys. Rev. B \textbf{94} 014425 (2016).
\bibitem{Freimuth} F. Freimuth, S. Bl\"{u}gel, Yu. Mokrousov, Phys. Rev. B \textbf{94} 144432 (2016).
\bibitem{Belotelov} V. I. Belotelov, A. K. Zvezdin, Phys. Rev. B \textbf{86}, 155133 (2012).
\bibitem{Kastler} A. Kastler, J. Phys. Radium \textbf{11}, 255 (1950).
\bibitem{Auzinsh} M. Auzinsh, R. Ferber, \emph{Optical Polarization of Molecules} (Cambridge University Press, Cambridge,
England, 1995).
\bibitem{Orient} \emph{Optical Orientation}, edited by F. Meyer, B.~P.~Zakharchenya (North-Holland, Amsterdam, 1984).
\bibitem{Kimel1} A. V. Kimel, F. Bentivegna, V. N. Gridnev, V. V. Pavlov,  R. V. Pisarev, Th. Rasing, Phys. Rev. B \textbf{63}, 235201 (2001).
\bibitem{Belykh} V. V. Belykh, E. Evers, D. R. Yakovlev, F. Fobbe, A. Greilich, M. Bayer, Phys. Rev. B \textbf{94}, 241202(R) (2016).
\bibitem{Wachter}P.~Wachter, \emph{Handbook on the Physics and Chemistry of Rare Earths}, Vol. 11, Eds. K.~A.~Gschneider, L.~R.~Eyring (North Holland, Amsterdam, 1979), p. 507.
\bibitem{Gunther} G.~G\"{u}ntherodt, Phys. Cond. Matter \textbf{18}, 37 (1974).
\bibitem{Nagaev}E.~L. Nagaev, \emph{Physics of Magnetic Semiconductors} (Mir, Moscow, 1983).
\bibitem{Lettieri}J.~Lettieri, V.~Vaithyanathan, S.~K.~Eah, J.~Stephens, V.~Sih, D.~D.~Awschalom, J.~Levy, D.~G.~Schlom, Appl. Phys. Lett. \textbf{83}, 975 (2003).
\bibitem{Schmehl}A.~Schmehl, V.~Vaithyanathan, A.~Herrnberger, S.~Thiel, C.~Richter, M.~Liberati, T.~Heeg, M.~R\"{o}ckerath, L.~F.~Kourkoutis, S.~M\"{u}hlbauer, P.~B\"{o}ni, D.~A.~Muller, Y.~Barash, J.~Schubert, Y.~Idzerda, J.~Mannhart, D.~G.~Schlom, Nature Mat. \textbf{6}, 882 (2007).
\bibitem{Averyanov} D. V. Averyanov, Yu. G. Sadofyev, A. M. Tokmachev, A. E. Primenko, I. A. Likhachev, V. G. Storchak, ACS Appl. Mater. Interfaces \textbf{7}, 6146 (2015).
\bibitem{Kaminski} B. Kaminski, M. Lafrentz, R. V. Pisarev, \emph{et al}., Phys. Rev. Lett. \textbf{103}, 057203 (2009).
\bibitem{Lafrentz} M. Lafrentz, D. Brunne, B. Kaminski, \emph{et al}., Phys. Rev. B \textbf{82}, 235206 (2010).
\bibitem{APL2010}K~Holldack, N.~Pontius, E.~Schierle, T.~Kachel, V.~Soltwisch1, R.~Mitzner, T.~Quast, G.~Springholz, E.~Weschke, Appl. Phys. Lett. \textbf{97}, 062502 (2010).
\bibitem{PRL2012} F.~Liu, T.~Makino, T.~Yamasaki, K.~Ueno, A.~Tsukazaki, T.~Fukumura, Y.~Kong, M.~Kawasaki, Phys. Rev. Lett. \textbf{108}, 257401 (2012)
\bibitem{SciRep2014} R.~R.~Subkhangulov, A.~B.~Henriques, P.~H.~O.~Rappl, E.~Abramof, Th.~Rasing, A.~V.~Kimel, Scientific Reports \textbf{4}, 4368 (2014).
\bibitem{NatureComm2015} M.~Matsubara, A.~Schroer, A.~Schmehl, A.~Melville, C.~Becher, \emph{et al}., Nature Comm. \textbf{6}, 6724 (2015).
\bibitem{Henriques1} A. B. Henriques, G. D. Galgano, P. H. O. Rappl, E. Abramof,
Phys. Rev. B \textbf{93}, 201201(R) (2016).
\bibitem{Henriques2} A. B. Henriques, A. R. Naupa, P. A. Usachev, V. V. Pavlov, P. H. O. Rappl, E. Abramof, Phys. Rev B \textbf{95}, 045205 (2017).
\bibitem{Heiss1} W. Heiss, R. Kirchschlager, G. Springholz, Z. Chen, M. Debnath, Y. Oka, Phys. Rev. B \textbf{70}, 035209 (2004).
\bibitem{Henriques3} A. B. Henriques, A. Wierts, M. A. Manfrini, et al., Phys. Rev. B \textbf{72}, 155337 (2005).
\bibitem{Heiss} W. Heiss, G. Prechtl, G. Springholz, Phys. Rev. B \textbf{63}, 165323 (2001).
\bibitem{Gunther1} G.~G\"{u}ntherodt, P. Wachter, D. M. Imboden, Phys. kondens. Materie \textbf{12}, 292 (1971).
\bibitem{Landau} L. D. Landau, E. M. Lifshitz, Phys. Z. Sowjetunion \textbf{8}, 101
(1935).
\bibitem{Gilbert} T. L. Gilbert, Ph.D. thesis, Illinois Institute of Technology,
Chicago, 1956.
\bibitem{Mondal} R. Mondal, M. Berritta, P. M. Oppeneer, Phys. Rev. B \textbf{94}, 144419 (2016).

\end{thebibliography}
\end{document}